# IoT Device Fingerprint using Deep Learning

Sandhya Aneja
*Faculty of Integrated Technologies*
*Universiti Brunei Darussalam*
Brunei Darussalam
sandhya.aneja@ubd.edu.bn

Nagender Aneja
*Institute of Applied Data Analytics*
*Universiti Brunei Darussalam*
Brunei Darussalam
nagender.aneja@ubd.edu.bn

Md Shohidul Islam
*Faculty of Integrated Technologies*
*Universiti Brunei Darussalam*
Brunei Darussalam
14b6057@ubd.edu.bn

*Abstract*—Device Fingerprinting (DFP) is the identification of a device without using its network or other assigned identities including IP address, Medium Access Control (MAC) address, or International Mobile Equipment Identity (IMEI) number. DFP identifies a device using information from the packets which the device uses to communicate over the network. Packets are received at a router and processed to extract the information. In this paper, we worked on the DFP using Inter Arrival Time (IAT). IAT is the time interval between the two consecutive packets received. This has been observed that the IAT is unique for a device because of different hardware and the software used for the device. The existing work on the DFP uses the statistical techniques to analyze the IAT and to further generate the information using which a device can be identified uniquely. This work presents a novel idea of DFP by plotting graphs of IAT for packets with each graph plotting 100 IATs and subsequently processing the resulting graphs for the identification of the device. This approach improves the efficiency to identify a device DFP due to achieved benchmark of the deep learning libraries in the image processing. We configured Raspberry Pi to work as a router and installed our packet sniffer application on the Raspberry Pi . The packet sniffer application captured the packet information from the connected devices in a log file. We connected two Apple devices iPad4 and iPhone 7 Plus to the router and created IAT graphs for these two devices. We used Convolution Neural Network (CNN) to identify the devices and observed the accuracy of 86.7%.

*Index Terms*—IoT, Fingerprint, Signature, Image Processing, Deep Learning, Raspberry Pi , Device identification

## I. INTRODUCTION

Packet sniffing is implemented using a packet sniffer, which is a software that deciphers, logs and investigates the network packet using packet capturing [1]. Packet capturing is a process of extracting information from the received packets of a device on a network. The packets are captured using packet filtering, which is a process of filtering the packets based on the type of packets, e.g. Internet Protocol (IP), Transmission Control Protocol (TCP), or User Datagram Protocol (UDP). Sniffing can be misused by a malicious user to destroy the security of the network by monitoring the network for a specific device address in a passive fashion. Some malicious users can even sniff to get confidential data (passwords, IDs, or other confidential information). Network administrators also use sniffing tool as an anti-intrusion tool to control the access of the network by only trusted devices. However, there are scenarios wherein a malicious device may spoof the IP address or MAC address to access the network.

DFP creates a unique signature for a device to identify the device without using its network identities like IP address or its physical identity like MAC address or IMEI number. DFP analyzes the information of the packets to extract IAT that is the time interval between consecutively received packets. In this paper, we worked on the DFP using IAT to create IAT fingerprint using deep learning. IAT is unique for each device due to the different hardware and software used for the device and thus is more useful in the wireless networks and Internet of Things (IoT) devices where communication is highly vulnerable due to identity issues.

This work presents a novel idea of DFP by plotting the graphs of IAT of packets. We plotted a number of graphs for consecutive packets of about 24 hours with each graph plotted the pattern of IAT for 100 packets. We connected two Apple devices iPad4 and iPhone7 Plus to Raspberry Pi . The packets were captured at the Raspberry Pi and plots generated by Python Programming Language. The IAT plots were subsequently processed for the identification of the device using Deep Learning. This approach using IAT graphs improved the efficiency of the DFP due to achieved benchmark of the deep learning libraries in the image processing. We achieved accuracy approaching near to 86.7%.

The organization of the paper is as follows: Section 2 defines problem statement, and Section 3 explains related work. The methodology and models used to generate the dataset for DFP for this paper is in Section 4. Experimental results and analysis are in Section 5 with the conclusion in Section 6.

## II. PROBLEM STATEMENT

Communicating devices use the TCP/IP model on the network. The devices can be identified on the network using IP address at Network Layer and MAC address at Data Link Layer. However, these identification schemes have been exploited by spoofing identities to gain access to the restricted resources.

IP spoofing is a technique through that a device can forge its network identity to access the network while MAC spoofing is a method for forging MAC address of a machine using a software emulating over a machine or device to bypass access control. For example, an ioctl system call can modify MAC address of a Network Interface Card (NIC) [2]. A scheme to extract the MAC address of the devices connected to Wi-Fi Access Point (AP) is discussed by Cunche et al. [3]. The fingerprint of the devices connected to AP is stored and replayed at another location assuming only the same device will connect to the forged AP. However, the authors have not presented any results on the presented idea.

Access control is used to distinguish between authorized and unauthorized devices competing for the network resources on a network using network identities like IP address and MAC address. For example, on some airports access to Wi-Fi connectivity is using the identity of the device. In many corporate offices, devices are also identified by their MAC addresses. However, an unauthorized device may spoof its network identity to get access to the corporate network.

Wireless Local Area Networks (LANs) are more prone to security issues than the wired LANs. The frames used to communicate on the wireless LANs are encrypted to make frames resistant to spoofing and replay but still, management and control frames are unencrypted under IEEE 802.11 standards for identity resolution through MAC address. This makes wireless LANs prone to spoofing and the denial of service attacks. The problem of spoofing is more severe in case of wireless multi-hop networks wherein any node without authorization may join the network.

Further, in sensor and IoT networks, sensors are used in open environments for communication. The number of IP based sensors for IoTs are in large quantity and are produced by various vendors sometime without considering the security aspects. Thus, unauthorized access to the open environments may lead to gain information from malicious devices to compromise the legitimate devices and upload its data with misleading information. Therefore, it is required to identify the devices without using their other identities.

DFP is a technique to identify a device using the information obtained for the targeted device. DFP is defined as a passive fingerprint which is obtained by observing the information sent by the target device in response to some action imposed by the intender. DFP is the possible solution for the IP spoofing and MAC spoofing problems in wired and wireless scenarios for the sensor networks, ad-hoc networks, IoT networks. In this paper, we present a method of DFP and distinguishing the devices using the DFP and Deep Learning.

### III. RELATED WORK

Uluagac et al. [4] presented DFP technique based on the statistical analysis of IAT for packets from wireless devices. The authors created the histograms where each bin in histogram represents the frequency of packets having IATs in a specified range. The generated set of histograms for each type of packet from a device is defined as its fingerprint (signature). The authors further used the similarity metric to identify the device from the created database of the signatures of the devices under study. The scheme works for the known and unknown devices. Known devices are those for which signature exists in the database while when the signature does not exist for the device, it is kept in the unknown category. The authors tested the system with various defaulted scenarios wherein the malicious user configure the device to generate the packets at the rate of another device, introduce delays to alter its communication pattern, vary the packet sizes, use some other protocol. The authors [4] concluded that the devices are different in their clock skews, system software and the applications installed on the devices and thus practically it is difficult to emulate another device and change the communication pattern. Thus, high accuracy in DFP schemes can help in access control and network management where security is an extremely important issue. The authors worked on passive device fingerprinting where the wireless devices were observed while uploading or downloading the data over the LAN using IPERF and PING applications.

Radhakrishnan et al. [5] extended Uluagac et al. [4] approach for active device fingerprinting by using ping application to communicate with the wireless devices on a live campus network. The authors found ping with 1400 bytes is better to be used since ping with 1400 bytes uses memory modules and incorporate signature of the device better into the arrival time.

Desmond et al. [6] presented DFP on wireless LANs through the timing analysis of the 802.11 probe request frames. The inter-arrival of frames are different on different client devices due to (i) operating system (ii) noise (iii) driver specific scanning process (iv) clock skew between machines. The authors observed that in wireless IEEE 802.11 standard LANs when probe frames are used there are inter-burst latencies between the frames. Moreover, inter-burst latencies were packed in groups rather than clustered around a centroid. These inter-burst latencies can be used to identify the device.

Miettinen et al. [7] discussed the methods of securing the IoT devices connected to an access point (AP) using the DFP approach. The authors extracted the set of 23 features from Link layer protocol (2- ARP/LLC), Network layer protocol (4-IP, ICMP, ICMPv6, EAPoL), Transport layer protocol (2-TCP, UDP), Application layer protocol (8-HTTP, HTTPS, DHCP, BOOTP, SSDP, DNS, MDNS, NTP), IP options (2- Padding, Router alert), Packet content (2- Size , Raw data), IP address (1- Destination IP address), Port class (2- source, destination) respectively from successive 12 packets resulting a feature set of size 23 X 12 = 276. Multi-class classifiers using random forest algorithm are used so that one classifier for one device type is trained. For a new device, it is assumed to be of one of a trained classifier which could save the relearning process in comparison to particular device identification wherein for every new device relearning is required. The authors did not use special packets to generate the traffic rather captured the communication by the device over the Internet. The accuracy of the experiment performed for 17 devices over a set of 27 devices was 95% and for the rest of 10 devices was 50% resulting in an average of 85%.

Kulin et al. [8] described publicly available wireless datasets of IAT and used the similar statistical approach of frequency of histogram bins for IAT in Radhakrishnan et al. [5], Uluagac et al. [4] for all considered datasets. Approach achieved the accuracy of 82% but precision and recall varied near to 99% using different machine learning algorithms for the proposed model.

Muhammad et al. [9] suggested to use their own devices by keeping access control on the devices using DFP since the DFP provide a unique identification of the devices. This is also applicable in the corporate offices where usually there is a restriction on to use only allowed devices.

Robyns et al. [10] proposed entropy-based per bit analysis of a frame for each mobile device. Authors categorized this scheme as non-cooperative MAC layer fingerprinting since the radio transmission by the mobile device captured by monitoring stations for fingerprinting does not require permission of a user. Non-cooperative MAC Layer frame bit analysis could be used by monitoring stations to track the location of devices by some adversary nodes which hampers the privacy of the user. The scheme proposes to analyze the bit pattern of the frame for variability and stability rate of each particular bit in the frame for all the devices. Randomization of MAC address is used to constraint the attackers to extract the MAC address from the frame. The accuracy of the scheme was found to vary between 67% to 80% for 50 to 100 devices while when the number of devices were increased, accuracy was found to vary 33% to 15%.

Kohno et al. [11] presented results of remotely fingerprinting the devices over the Internet. They measured clock skews of the devices to fingerprint the device even when devices changed their IP addresses or shifted in time assuming that the trace of devices is available using tcpdump. The difference of time-stamp in the packets is bounded by the difference of arrival time of the packets. They measured the clock skew of the devices by solving for the rate of change of the time-stamp in packets.

Xu et al. [2] presented opportunities and challenges on DFP approaches for wired and wireless networks with a taxonomy of features used for DFP from various layers of the protocol stack. Mainly features are studied from the physical layer and MAC Layer varying from transmission time, inter-arrival time, clock skewness, per bit analysis of frames for amplitude, frequency, SSID etc. IAT and transmission time features are found to be promising parameters for accuracy. The main challenge in the approach is that it uses synthetic or simulated data. Public data for a large number of devices for all the scenarios are not available.

Maurice et al. [12] presented a co-operative DFP using IEEE probe request frames for similar devices connected to Access Point (AP). DFP uses traffic analysis of the devices over the network which is assumed to follow a different flow for different devices. The accuracy of DFP does not show promising results in case of similar devices. Authors suggested co-operating the similar devices by modifying some traffic attribute so to result in unique signature for each device.

Cunche [13] presented linking of wireless devices using probe request messages over the preferred APs. In passive scanning, AP sends probe request messages while in active scanning device sends probe request messages to discover the available list of wireless devices. Devices broadcast the set of preferred networks in the probe request messages in passive scanning. This is the vulnerability of IEEE 802.11 standard which can be used to identify the people linking to each other with the similar list of preferred networks.

François et al. [14] proposed to use DFP for enforcing authentication and generating access control rules automatically based on the observed behavior of the device. The authors called this approach as behavioral fingerprinting. For example, a device misusing the network resources can be enforced to reconnect the network based on its behavioral fingerprint.

Sun et al. [15] used DFP for localization of devices connected to the Wi-Fi AP in indoor as well as outdoor experiment setups. In their approach, the first database of DFP is maintained in the experiment and then on a geometrical set up of devices

## IV. DEVICE FINGERPRINTING USING DEEP LEARNING

We propose the following four models to capture packet information on the router based on the layers and methodology used.

### A. DFP Model 1

*DFP Model 1* explains a scenario where all devices are connected to a router by Wi-Fi and using probe frames.

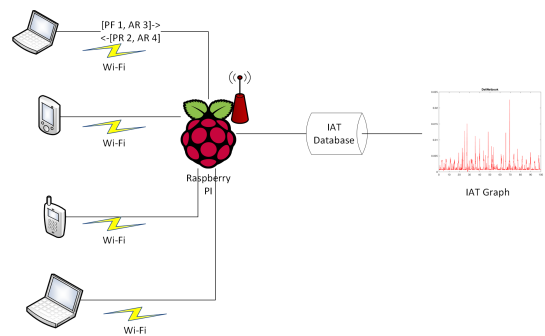

Fig. 1. DFP Model 1 using Probe Frames

Figure 1 shows Raspberry Pi as a router connected to various user devices and uses probe frames for IAT database [6]. A sniffing application can be utilized to capture the frames used for communication by the client devices with the router. The router first broadcast the probe frames to search the devices in the range within a specific period and then the devices respond to the router with reply frames periodically. Source MAC address, destination MAC address, SSID, and arrival time of response frames from different devices can be used for DFP

### B. DFP Model 2

*DFP Model 2* explains a scenario where all devices are connected to a router by Wi-Fi and using Ping Packets.

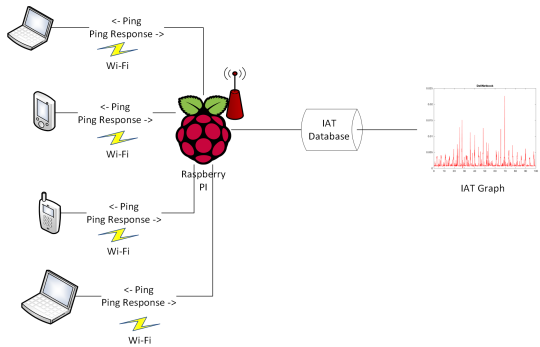

Fig. 2. DFP Model 2 using Ping Packets

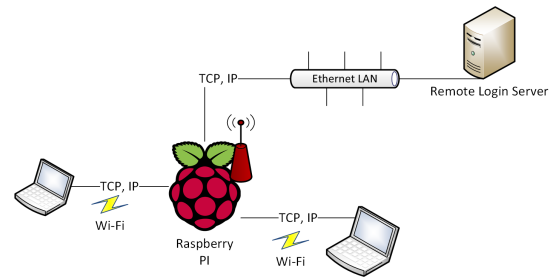

Fig. 4. DFP Model 4 using TCP or IP packets

Figure 2 shows Raspberry Pi as a router connected to the various users with sniffing and ping application capturing the ping packets used for communication by the devices with the router. A ping application [5, 4] on the router can be used to get the IAT for different devices on the network. IAT of ping packets can be compared to check the accuracy of schemes based on the different types of packets used among the devices.

*C. DFP Model 3*

*DFP Model 3* explains a scenario where all devices are connected to a router by Wi-Fi and using UDP, Address Resolution Protocol (ARP) or Internet Control Messaging Protocol (ICMP) packets.

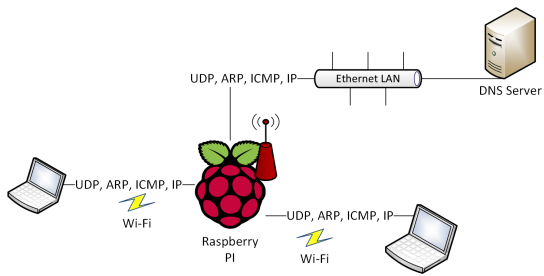

Fig. 3. DFP Model 3 using UDP, ARP or ICMP packets

Figure 3 shows router connected to various user devices using IEEE 802.11 LAN standards. The sniffing application captures the UDP, ARP or ICMP packets used for communication by the devices with router [5, 4]. The application on the router can be used to get the arrival time for different devices on the network.

*D. DFP Model 4*

*DFP Model 4* explains a scenario where all devices are connected to a router by Wi-Fi and using TCP or IP packets.

Figure 4 shows router connected to various user devices using IEEE 802.11 LAN standards. The sniffing application captures the TCP or IP packets used for communication by the devices with router [5, 4]. The application on the router can be used to get the arrival time of packets for different devices on the network.

*E. Convolution Deep Network Learning Model*

The objective of this paper is using CNN to recognize a device based on the IAT graphs. As discussed earlier, the graphs of the IAT were plotted. The plotted graphs were used by CNN as image classification to analyze if the algorithm can distinguish between the devices solely based on IAT plots without considering any other information. As we collected IAT data from two devices iPad4 and iPhone7 Plus and applied CNN to distinguish the devices using IAT.

Many researchers have proposed that deep learning is appropriate when there is a significant amount of data. However, this statement is little misrepresenting. Deep learning algorithm functions admirably when it can learn features, and that is possible when the training data is extensive. However, CNNs are the excellent model for image classification even with limited training data. Since we are taking few examples of IAT plots, we augmented the data by a number of random transformations so that the model doesn't see the same picture twice. This also prevents over-fitting and generalize the model.

A shear transformation is a linear transformation that displaces each point in a fixed direction. We performed the shear mapping in the vertical direction by the amount of 0.2 times of the distance from the edge of the image. This number can be changed in different experiments. Zoom transformation performs zooming in or out. We performed zooming on the IAT plots by 0.2. We performed horizontal flipping that flips the image w.r.t the vertical axis. We generated 636 IAT graphs for iPad4 and 608 graphs for iPhone 7 Plus.

First, we declared an input layer to process the input images. The input shape parameter of the input layer should be the shape of each image i.e. (depth, width, height). We converted all images of size 150x150, thus the input shape parameter is (3, 150, 150). The input shape is required only in the input layer. We used 32 convolution filters with kernel size of 3 rows and 3 columns moving window. We considered Rectified Linear Unit (ReLU) as activation function. Next, we added a 2D max-pooling layer with the size of pooling as 2 in x-direction and 2 in the y-direction. The max-pooling reduces the number of parameters by sliding a 2x2 pooling filter by taking the maximum of the 4 values.

Next, we added two convolution and max-pooling layers with 32 filters and 64 filters with kernel size 3x3, ReLU and max-pooling and stride size 2x2.

We considered the batch size of 16. The batch size refers to a number of samples that are propagated through the network. It means training images will be processed in the batches of 16 images. Setting batch size less is important that it requires less memory and the training is fast since the weights are updated after each propagation. However, a too much smaller value of batch size may result in a less accurate estimate of the gradient.

We considered epochs size as 50. One epoch refers to one forward pass and one backward pass of all training images. Thus, in case of an example with 500 training images and batch size 250, it will take 2 iterations to complete 1 epoch.

## V. Implementation and Results

We set up the system for the experiment of our paper using two Apple devices iPhone 7 Plus and iPad 4. We deploy our sniffer application over the router running Raspbian OS configured as a Wi-Fi hotspot and connected to wired LAN running Ethernet services. Packet sniffing application with filters for different types of packets logged arrival time with features, e.g. IP addresses, MAC addresses and Port addresses of the respective devices connected to the router shown in the Figures 1 - 4. We propose four types of DFP models for the devices using ping application and probe frames of IEEE 802.11 standard. Ping application deployed over the router reads the IP addresses of the devices logged by the sniffer application and record the IAT of packets over the network. We generated 636 IAT graphs for iPad4 and 608 graphs for iPhone 7 Plus. The graphs were randomly divided between training and validation data set in the ratio of 80% and 20%. The following table explains the data set.

TABLE I
Training and Validation Data Set

|  | Training | Validation | Total |
|---|---|---|---|
| iPad4 | 509 | 127 | 636 |
| iPhone 7 Plus | 486 | 122 | 608 |
| Total | 995 | 249 | 1244 |

We achieved the accuracy of 86.7% using this approach.

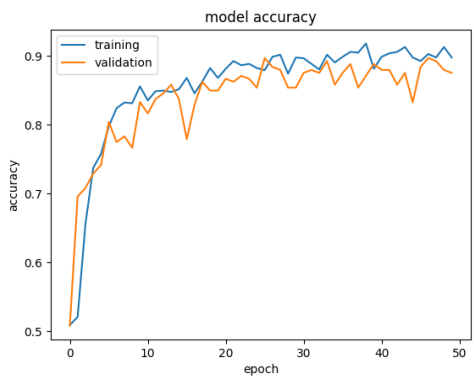

Fig. 5. Model Accuracy

Figure 5 shows progress in the accuracy as increasing epochs. As we can see the number of epochs that we selected as 50 were optimum to have the desired accuracy.

Figure 6 shows as for how the loss in the model is reduced as epochs increased. A loss is also called mean squared error, and lower the value of the loss, better the predictions.

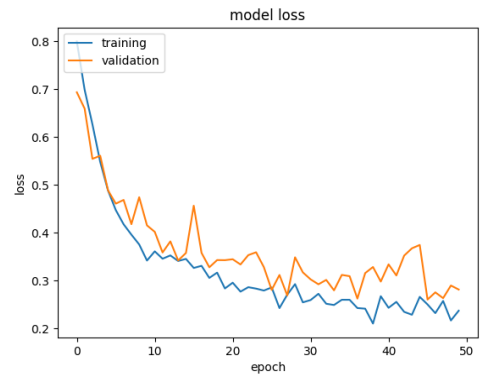

Fig. 6. Model Loss

## VI. Conclusion

In this paper, we focused on using Packet Capture (PCAP) library routines, setting up Raspberry Pi as the router, developing a new DFP approach and then setting the experiment to test the approach on our dataset captured real-time using Raspberry Pi .

Many researchers have worked on the problem of DFP using machine learning, however, we are the first one to propose deep learning for DFP. The previous approaches are complex in terms of granularity of choosing the feature set which further includes approximations. Also, a few researchers have presented their results using DFP based on a particular packet type which is not practical to use due to being applicable for that packet type only. We used packets flowing from the device to router for DFP and achieved the desired accuracy.

The approach is useful for both wireless and wired scenario as an alternative for maintaining the identities of the devices on the network. In the case of wired scenario, it helps to monitor the devices, and in case of wireless scenario, it presents another alternative for identity. Although DFP replay has been used for stealing the identity and replaying, however, on replaying it is used on another hardware and thus the signature changes. Thus, DFP can help to detect malicious users if devices signature is stored and matched before allowing users to connect the network.

Our scheme of deep learning is not computing intensive since it uses mostly time series graphs which are sparse matrices and thus computationally fast. Model completes in few minutes matching the need for practical scenarios. Our model using two Apple devices achieved the accuracy of 86.7% that can further be improved by having a better router. Since Raspberry Pi has limited memory and is slow when multiple devices are connected to Raspberry Pi as the router.

A device using hardware of Raspberry Pi takes more time to capture the sufficient number of packets to generate a signature of 8-10 devices. Moreover signature needs to be generated on cloud due to a slow computation of embedded devices. Apart from the experiment limitations, our experiment data was legit which has shown a highly accurate result.

The proposed approach is useful for IoT devices which are used in an open scenario for data collection. Hackers may spoof IMEI number also; however, DFP can still help in detecting the device on a cellular network. DFP can also help in identifying the devices in co-operation based Multi-hop wireless networks. We also observed inter-burst latencies in the IAT data and study can be conducted to capture the outliers of the data set and further using to see if the plots of outliers help in detecting the device. It is also suggested to configure the access point to prevent replicating the DFP when it detects different hardware.